# Numerical Investigation of Wave Scattering in Granular Media: Grain-Scale Inversion and the Role of Boundary Effects[1]

Ning Liu [1] (✉), Wen-Tao Hu [1]

[1.] College of Mechanical and Electrical Engineering, Beijing University of Chemical Technology, Beijing, 100029, China
nicolaliu@buaa.edu.cn

**Abstract.** Seismic coda waves, once dismissed as mere noise, are now recognized as critical signatures of wave scattering in seismograms. In 1969, Aki proposed that these waves originate from small-scale heterogeneities within the Earth's interior, sparking a new research focused on their interpretation and application. Subsequent studies have demonstrated that coda waves carry rich information about subsurface heterogeneity, making investigating scattering phenomena essential for probing complex geological structures. Wave propagation and scattering in unconsolidated granular media are particularly relevant for understanding the seismic behavior of planetary regoliths, such as those found on the Moon and Mars. The energy-based radiative transfer equation (RTE) offers a theoretical framework to quantitatively relate scattering characteristics to microstructural properties, including grain size, coordination number, and porosity. However, the RTE assumes an infinite medium, an assumption rarely satisfied in realistic scenarios, underscoring the need to evaluate how boundary effects influence wave scattering and inversion accuracy.

[1]This study leverages the discrete element method (DEM) to simulate elastic wave propagation and scattering in granular media. Unlike conventional numerical approaches, DEM explicitly models grain-scale interactions, including contact forces and dynamic behavior, enabling detailed resolution of wavefield features shaped by microstructural heterogeneity and boundary geometry. Using the RTE framework, we invert scattered wave energy to estimate microstructural parameters and assess the effects of absorbing (infinite) and rigid (finite) boundary conditions on inversion performance. The results show that DEM effectively reproduces wavefields in granular media and that RTE-based inversion is a feasible approach for retrieving grain-scale properties. However, boundary reflections significantly distort the wavefield, leading to substantial errors in the inversion outcomes. This research provides new insights into wave scattering in granular materials and offers theoretical guidance for

designing and interpreting seismic experiments in planetary regolith environments.

# 1    Introduction

Seismic coda waves are the trailing part of a seismogram that occurs after the direct arrivals of seismic waves, long understood as the result of multiple scattering caused by variations in the Earth's crust. In 1969, Aki[1] first proposed that the coda of local earthquakes includes waves scattered multiple times by small-scale inhomogeneities in the medium. Each subsequent coda arrival follows a long and complex path through the Earth, which means that the scattering geometry influences the shape and duration of the coda. More significant heterogeneity in the crust leads to longer and more pronounced coda signals. Laboratory and numerical studies have confirmed that coda waves sample the entire medium multiple times, making them much more sensitive to small-scale changes than the first arrivals of waves. Building on these findings, researchers have also shown that the envelopes of coda waves contain information about the seismic source and the scattering characteristics of the medium. Today, coda waves are broadly recognized as emergent energy produced by waves bouncing off random heterogeneities rather than being classified purely as direct body waves[1,2].

Scattering caused by subsurface heterogeneity significantly impacts the interpretation of seismic data. The decay and shape of coda envelopes provide valuable information regarding the density, size, and distribution of scatterers. Researchers commonly use the radiative transfer equation (RTE) and related diffusion models to connect coda attenuation with volume-averaged scattering properties. However, although these models offer insights into the average size of scatterers and the mean free path, they often rely on idealized assumptions. For example, they frequently assume infinite, homogeneous media and ignore boundary conditions like free surfaces and finite layer thicknesses that are common in real-world geological settings[3].

Recent advancements have shown that coda waves provide more information than just attenuation data. Techniques such as coda-wave interferometry and waveform inversion leverage the sensitivity of coda waves to detect minute changes throughout the entire medium[4]. For instance, variations in velocity or evolving scatterers can be identified more effectively using coda arrivals than direct wave phases. Both laboratory and field studies have successfully inverted source parameters and small-scale heterogeneities using coda-based analyses. Thus, rather than being mere noise, coda waves constitute a diffuse wavefield rich in information, facilitating grain-scale inversions and continuous monitoring of subsurface changes.

However, most existing studies have concentrated on crystalline, sedimentary, or dense granular media[5,6]. In contrast, loose granular materials, such as planetary regoliths, remain relatively unexplored in this context. Regolith, the unconsolidated surface layer on bodies like the Moon and Mars, displays significant granular heterogeneity. Apollo-era seismic data have suggested intense scattering within the lunar crust, and numerical models indicate that accurately reproducing coda waveforms necessitates pervasive near-surface heterogeneities. For example, research by Onodera et al.[7,8] has shown that models incorporating widespread heterogeneity in the upper 10-20 km effectively matched Apollo 12 impact codas. These models indicate extremely low P/S velocity ratios, consistent with dry, porous regolith. Similarly, seismic data from the InSight lander on Mars have revealed a loose, low-velocity regolith layer just a few meters thick, likely contributing to strong near-surface scattering[9]. These environments challenge the applicability of traditional RTE models designed for solid rock media.

Researchers commonly employ the energy-based RTE framework to extract scattering information from coda waveforms. This framework relates coda energy's spatial and temporal decay to average scattering parameters, such as cross-section and mean free path. However, conventional implementations of the RTE typically assume an infinite, homogeneous medium, neglecting boundary conditions like free surfaces or stratification. When applied to granular materials, these assumptions can lead to inaccuracies, where the layer thickness may be comparable to the scattering mean free path.

To tackle these challenges, we utilize the discrete element method (DEM), a numerical approach initially developed by Cundall and Strack[10] for modeling granular and fractured materials. DEM also provides crucial advantages for simulating strongly continuous systems and allows for seamless transitions between solid and particulate phases[11]. Each grain is explicitly represented, with its dynamic interactions, e.g., contacts, collisions, and friction, fully resolved. DEM has been effectively applied in geophysics to model granular flows and their associated seismic emissions. For instance, Arran et al.[12] used DEM to simulate "slidequake" signals generated by landslides, exploiting the method's ability to access particle-scale information beyond experimental methods' reach. Additionally, DEM can resolve elastic wave scattering caused by individual grains in static granular systems.

Importantly, DEM allows for explicit modeling of boundaries and free surfaces. This study applies DEM to simulate elastic wave propagation in bounded granular media and to assess the impact of boundary effects on coda wave characteristics. We employ a perfectly matched layer (PML) to approximate an infinite medium as an absorbing boundary condition. By comparing DEM simulations using PML with those in finite samples, we evaluate the significance of boundary reflections and their effects on the accuracy of grain-scale inversions. In this way, we bridge the gap between idealized RTE models and realistic granular materials, providing a first-principles analysis of coda scattering theory in complex, bounded granular media.

## 2 Methodology

### 2.1 Radiative Transfer Equation

Radiative transfer equation (RTE) was first introduced in astrophysics as a phenomenological model to describe how light energy travels through the atmosphere[13]. Later, a more rigorous derivation based on the wave equation was developed, providing a solid theoretical foundation for the RTE. In seismology, Wu[14] and Hoshiba[15] adapted the RTE framework to explain the generation of seismic coda waves, attributing these waves to the scattering of seismic energy by small-scale heterogeneities within the Earth's interior.

In many applications, scattering is often treated as isotropic and acoustic, which leads to the neglect of the whole vector nature of elastic wave propagation, particularly the conversions between compressional (P) and shear (S) waves[16]. Despite this simplification, the RTE remains a valuable tool for analyzing the incoherent energy transport associated with high-frequency seismic wave envelopes in heterogeneous media[17].

This study focuses on isotropic multiple scattering from randomly distributed scatterers to model the propagation of elastic wave energy in two-dimensional random inhomogeneous media. Equation (1) expresses the general form of the RTE for this system in terms of the spatial coordinate vector $\mathbf{r}$:

$$E(\mathbf{r},t) = WG(\mathbf{r},t) + \xi \iiint G(\mathbf{r}-\mathbf{r}',t-t') \times E(\mathbf{r}',t') \mathrm{d}\mathbf{r}' \mathrm{d}t'. \tag{1}$$

Equation (2) shows an analytical solution to the RTE under simplified assumptions for the two-dimensional case[18]:

$$E(\mathbf{r},t) = W \frac{\mathrm{e}^{-\frac{vt}{l_s}}}{2\pi v r} \delta(t-\frac{r}{v}) + W \frac{\mathrm{e}^{\frac{\sqrt{v^2t^2-r^2}-vt}{l_s}}}{2\pi l_s \sqrt{v^2t^2-r^2}} H(t-\frac{r}{v}). \tag{2}$$

This analytical expression is utilized in our study to perform grain-scale inversion, allowing us to estimate the characteristic length scale of heterogeneity within each model domain. We introduce dimensionless variables for both time and transmission distance to facilitate our analysis, resulting in a dimensionless form of the energy density equation:

$$\varepsilon(\gamma,\tau) = \frac{e^{-\tau}}{2\pi\rho} \delta(\tau-\gamma) + \frac{1}{2\pi\sqrt{\tau^2-\gamma^2}} e^{\sqrt{\tau^2-\gamma^2}-\tau} H(\tau-\gamma) \tag{3}$$

$$\gamma = \frac{r}{l_s}, \tau = \frac{Vt}{l_s}, \varepsilon(r,\tau) = \frac{El_s^2}{W}. \tag{4}$$

Figure 1 presents the corresponding numerical results.

(a)                                        (b)

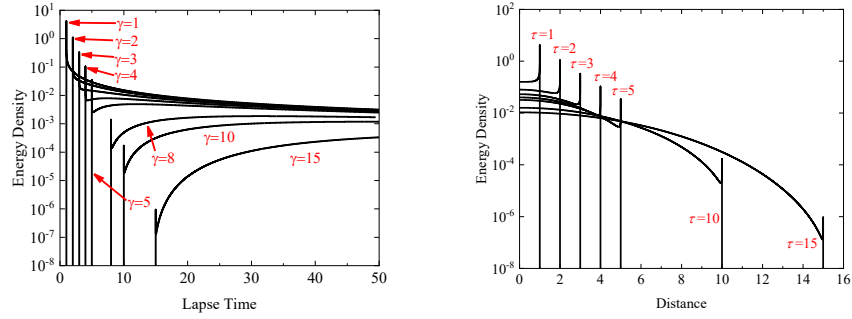

**Fig. 1.** Theoretical solutions of RTEs for (a) dimensionless time and (b) transmission distance

## 2.2 Discrete Element Method

The Discrete Element Method (DEM) models materials by dividing them into rigid elements with simple shapes, typically using circular (in 2D) or spherical (in 3D) forms due to their efficient contact detection capabilities. A typical DEM analysis involves three main steps: Computes forces at contact points using interaction laws; determines particle velocities and displacements by integrating Newton's second law; identifies new contacts and removes broken ones.

DEM treats interactions as a dynamic process, applying Newton's second law to calculate accelerations based on net forces, including gravity and internal contact forces. These accelerations are then integrated over time to update velocities and displacements. Contact forces are calculated using a force-displacement law, incorporating stiffness, damping, and friction models. Due to its efficiency and explicit nature, the central difference method is commonly used for time integration in DEM[11].

**Theoretical Wave Velocity of DEM.** For the discrete element model with triangular arrangement, the element's average density corresponds to the circumscribed rectangle of particles of equal mass, so the average density is as follows:

$$\rho_{\mathrm{m}} = \frac{\pi R^2 \rho}{4R^2} = \frac{\pi}{4}\rho. \tag{5}$$

The average area of the basic unit in the model can be calculated based on the basic unit shown in Fig. 2. The connection between the elements is a normal spring with stiffness of $K_{\mathrm{n}}$, an original length of $L$, and a spacing between the elements of $d$. If $S$ represents the rectangular area shown in Fig.2 and $N$ represents the number of effective elements in the zone, then we can calculate $S/N = (\sqrt{3}/4)d^2$. According to the derivation of Hoover et al.[19], assuming $\delta L/L \equiv \Delta$, the energy at "macro" level can be written as follows,

$$\Phi = \left[\frac{1}{2}\lambda(2\Delta)^2 + \frac{1}{4}\eta[(2\Delta)^2 + (2\Delta)^2]\right]S = 2\Delta^2(\lambda + \mu)S. \tag{6}$$

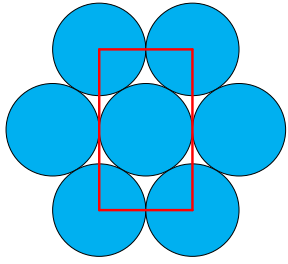

**Fig. 2.** triangular arrangement model

The deformation of each spring is $d\Delta$, so the "microscopic" energy can be written as:

$$\Phi = (3N)(\frac{1}{2}Kd^2\Delta^2)S = \sqrt{3}K\Delta^2 S.  \qquad (7)$$

According to the relationship between macroscopic and microscopic energy, combining Eqs. (6) and (7), we obtain as follows:

$$\lambda + \mu = \frac{\sqrt{3}}{2}K_n,  \qquad (8)$$

$$\mu = \lambda = \frac{\sqrt{3}}{4}K_n.  \qquad (9)$$

Substituting Eqs. (8) and (9) into the wave equations, we can obtain the wave velocities:

$$v_P = D\sqrt{\frac{9K_n}{8m}}, v_S = D\sqrt{\frac{3K_n}{8m}}  \qquad (10)$$

where $v_P$, $v_S$, $D$, $K_n$ and $m$ present the compressional wave velocity, the shear wave velocity, the diameter, the connection stiffness, and the mass of the element, respectively.

**Absorbing Boundary Condition.** The implementation starts by recognizing that the RTE is designed for an effectively infinite medium. As a result, numerical simulations must truncate the domain and incorporate absorbing layers to simulate the infinite boundary conditions. Typically, this is accomplished using perfectly matched layers (PML), artificial lossy layers at the boundary that allow outgoing waves of any frequency and angle to pass through without reflection. Berenger's PML[20], serves as the standard for this method and has been widely adapted for applications in acoustics and elasto-dynamics. Hastings et al.[21] demonstrated that incorporating a PML in an elastic wave simulation leads to excellent broadband absorption and stable performance, even under oblique incidence, surpassing traditional boundary schemes.

In a DEM simulation, we mimic the PML by adding impedance-matched forces on the boundary elements,

$$F = -\frac{2}{3}\pi R^2 \rho \sqrt{\frac{6K_n}{2\pi R\rho}}v(t),  \qquad (11)$$

where $R$, $K_n$ and $v(t)$ are the radius, the normal stiffness, and the velocity function of time $t$ of the elements. Concretely, each boundary element that comes into contact with an absorbing "wall" is assigned a normal force consisting of two components: a spring term related to the normal contact stiffness and a dashpot term proportional to the element's normal velocity. This approach effectively creates a soft and dissipative contact between the element and the wall, allowing incident acoustic energy to be trapped and dissipated at the boundary.

## 3    Numerical Results

### 3.1    Model Setups

This study utilizes a linear elastic contact model of DEM to simulate wave propagation in granular media. Table 1 details the input parameters for the simulations. Based on these parameters, the theoretical P-wave and S-wave velocities are 1200 m/s and 692 m/s, respectively. We use a Ricker wavelet with a central frequency of 400 kHz as the point source excitation, a common choice in seismic modeling.

The selection of a 400 kHz central frequency ensures that the dimensionless wavenumber satisfies the condition $ka \approx 1$. This condition guarantees that wave propagation occurs within the Mie scattering regime, characterized by complex scattering phenomena due to the comparable size of the scatterers to the wavelength. Operating within this regime allows for a detailed examination of wave scattering behaviors relevant to granular materials.

**Table 1.** Input parameters of discrete element models.

| Parameter | Value |
|---|---|
| Diameter $D$ (mm) | 0.70 |
| Density $\rho$ (kg/m$^3$) | 2650.00 |
| Normal stiffness $K_n$ (N/m) | $3.00*10^9$ |
| Shear stiffness $K_s$ (N/m) | 0 |
| Numerical damping | 0 |
| Viscous damping | 0 |

### 3.2    Validation and Calibration

Figure 3 shows the model and receiver arrangement, where the red elements represent boundary elements, and $L$ is the measurement distance of each numerical simulation. The receiver in the $x$ direction of the excitation source is Receiver 1, and the other measurement points are defined as Receivers 2 to 8 in clockwise order.

(a)                                              (b)

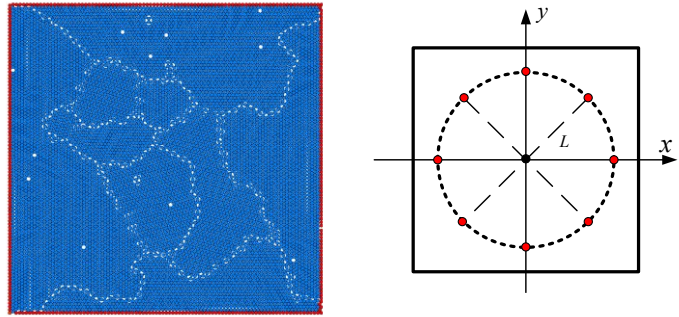

**Fig. 3.** Model and receiver arrangement

Figure 4 (a) and (b) show an obvious reflection wave in the rigid boundary model, while in the PML absorption boundary model, the reflection wave exists but is very weak. Figure 4 (c) shows the velocity variation curve of Receiver 1 over time. The PML absorption reflection wave absorption efficiency can reach 92.11%, proving that the PML model can cut off the reflection wave well. These results show that the PML absorption boundary setting can approximate the infinite medium.

(a) Rigid boundary      (b) PML boundary      (c) Velocity of Receiver 1

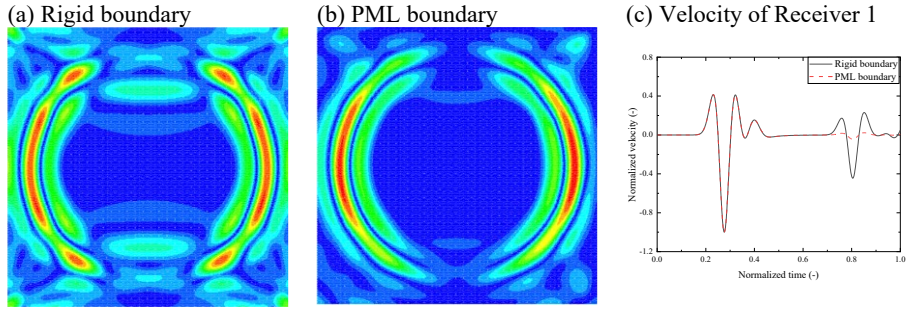

**Fig. 4.** wavefield of DEM models with (a) rigid boundary and (b) absorption boundary at *t*=0.3s, and Velocity of Receiver 1

### 3.3 The roles of Boundary Effect

**Absorption Boundary.** Figure 5 presents the numerical results for two measurement distances: 7.0 mm and 14.0 mm. Panels (a) and (b) show the velocity-time response for Receiver 1. After determining the energy density curve for the specified measurement distances, we normalize the energy envelope curve (illustrated in grey) based on a random time window from Receiver 1. We smooth these curves to enhance clarity, resulting in the black curve displayed in panels (c) and (d).

We fit the scattering free path using the analytical solution of the energy radiation transfer equation. This approach allows us to invert the scattering-free path of the model and characterize the grain-scale within the discrete element models. For measurement distances of 7.0 mm and 14.0 mm, the fitted scattering-free paths are 0.68 mm and 0.72 mm, respectively, with errors less than 3.00% within an acceptable range. These results indicate that the discrete element method suits wavefield forward simulation and grain-scale inversion studies of granular media.

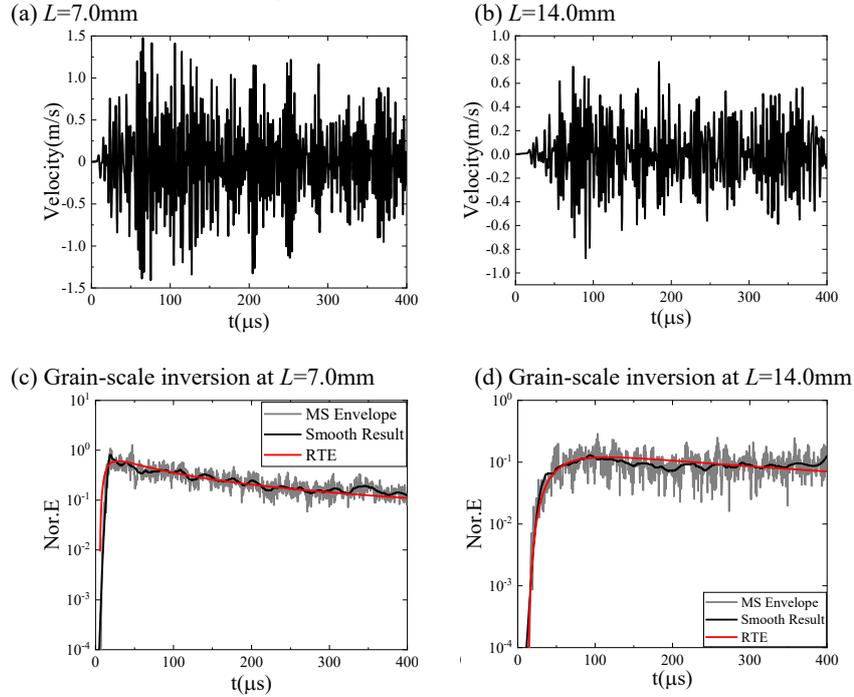

**Fig. 5.** Velocity curves of Receiver 1 and grain-scale inversion

**Rigid Boundary.** Figure 6 illustrates a discrete element model featuring a rigid boundary, with a height of 10.8 mm and a width of 30.0 mm. The blue element at the center represents the excitation source, while the green elements serve as the receivers. The distances from both receivers to the source are equal.

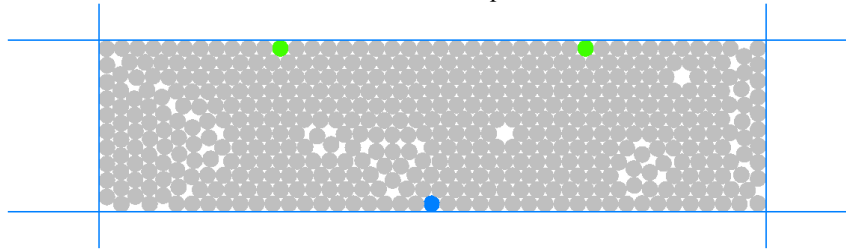

**Fig. 6.** Rigid boundary model with an excitation and receivers

Figure 7 (a) illustrates the inversion result when the measurement distance is 9 mm. The velocity curve presented in this figure represents the velocity response from one of the receivers. The MS envelope energy curve shown in Fig. 7 (b) is then theoretically fitted to perform grain-scale inversion. The normalization process involves selecting the energy from a specific time window to normalize the entire dataset and smoothing the data to achieve the smooth black curve shown in Figure 7(c). According to the RTE, the red curve represents the theoretical solution for this smooth curve fitting. The calculated scattering free path is 0.66 mm, with an error of 5.71%. This result indicates that boundary conditions significantly influence the results, and their effects are unavoidable during actual inversion.

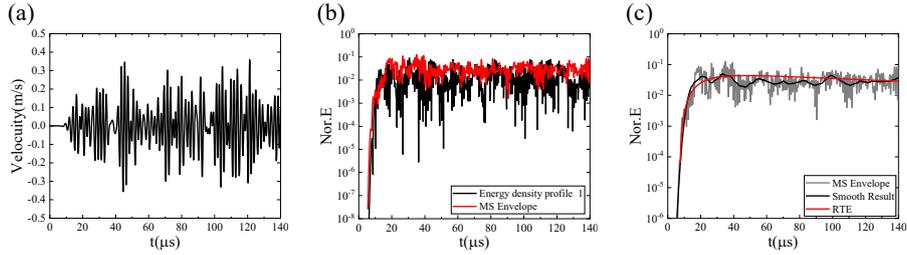

**Fig. 7.** Grain-scale inversion results with *L*=9.00 mm

## 4    Conclusions

This study demonstrates the utility of DEM for simulating wave scattering in granular media and highlights the feasibility of using RTE-based inversion to estimate microstructural properties. Crucially, we show that boundary effects must be explicitly accounted for in numerical models to avoid systematic errors in inversion results. These findings provide foundational knowledge for designing seismic experiments and interpreting data in planetary regolith environments. The insights gained here are particularly relevant for future seismic missions to the Moon and Mars, where accurate subsurface characterization depends on robust inversion strategies adaptable to constrained observational conditions.